# Active oversight and quality control in standard Bayesian optimization for autonomous experiments

*Sumner B. Harris,[1] Rama Vasudevan,[1] Yongtao Liu[1]* *

[1] Center for Nanophase Materials Sciences, Oak Ridge National Laboratory, Oak Ridge, Tennessee 37831, United States.

*Correspondence should be addressed to: liuy3@ornl.gov

**Abstract**

The fusion of experimental automation and machine learning has catalyzed a new era in materials research, prominently featuring Gaussian Process Bayesian Optimization (GPBO) driven autonomous experiments navigating complex experimental conditions for accelerated scientific discovery. In traditional GPBO-driven experiments, a predefined scalarizer function is often required to preprocess the experimental data, transforming non-scalar raw data into scalar descriptors for GP training. However, such predefined scalarizer functions have limitations, which likely fail to accommodate the diversity and complexity of real-world experimental data, potentially skewing experimental outcomes. Thus, oversight and quality control are necessitated over the process to avoid GPBO from being misled by low quality scalarizers. To address the limitation, we introduce a Dual-GP approach that enhances traditional GPBO by adding a secondary surrogate model to dynamically constrain the experimental space based on real-time assessments of the raw experimental data. This Dual-GP approach enhances the optimization efficiency of traditional GPBO by isolating more promising space for BO sampling and more valuable experimental data for primary GP training. We also incorporate a flexible, human-in-the-loop intervention method in the Dual-GP workflow to adjust for unanticipated results. We demonstrate the effectiveness of the Dual-GP model with synthetic model data and implement this approach in autonomous pulsed laser deposition experimental data. This Dual-GP approach has broad applicability in diverse GPBO-driven experimental settings, providing a more adaptable and precise framework for refining autonomous experimentation for more efficient optimization.

**Introduction**

The combination of experimental automation and machine learning techniques has ushered in a transformative era of autonomous experimentation, revolutionizing materials research by accelerating scientific discovery through high-throughput processes and data-driven decision-making.[1] Bayesian Optimization[2] (BO) plays a pivotal role in autonomous experiments for efficient optimization of target (objective) properties and exploration across extensive experimental conditions.[3,4] BO starts by making a statistical approximation of the unknown objective function in the experimental space, called a surrogate model, based on results from previously conducted experiments. The surrogate model for BO could be a random forest[5] or neural network[6,7] but is most commonly a Gaussian Process (GP)[8] due to its non-parametric nature and built-in uncertainty quantification[9,10]. Once the surrogate model is constructed, a variety of acquisition functions[11] can be calculated to determine the next set of experimental conditions that potentially reduce the surrogate's uncertainty, approach the global optimum, or maximize understanding of the system. Through this process, GPBO efficiently navigates vast experimental spaces, optimizing a target property or enhancing understanding with a minimum number of costly experiments.

GPBO has demonstrated applications in many materials science areas from theoretical predictions and materials design, to materials synthesis and characterization. GPBO has been used for the prediction of crystal structures[12,13] and for theoretical design of materials[14-16]. Autonomous synthesis methods that employ GPBO to efficiently optimize a target property vary from carbon nanotube growth[17-19], chemical synthesis[20,21], physical vapor deposition[22-24], and additive manufacturing[25-28]. It has enabled autonomous exploration and discovery in piezoresponse force microscopy (PFM)[29-31], scanning tunneling microscopy[32,33], and scanning transmission electron microscopy[34].

In the examples discussed above, a GP is utilized to map the relationship between inputs (e.g., chemical compositions, synthesis parameters, and characterization parameters) and outputs (e.g., target material properties). When target properties are quantifiable through scalar measurements, the scalar descriptors of target properties can be directly employed with the corresponding input parameters for GP training. However, in many real-world experiments, material properties are characterized from non-scalar data like spectroscopy, images, hyperspectral images, or higher dimensional and multi-modal data. This necessitates the use of *"scalarizer*

*functions"* that derive meaningful *scalar descriptors* from the *non-scalar raw data*; subsequently, these *scalar descriptors*, instead of the raw data, are employed along with the corresponding measurement or synthesis parameters for GP training.

Scalarizer functions can vary from simple functions like peak identification[35] to custom algorithms for specific physical attributes like the coercive field from polarization-voltage hysteresis loops in PFM[29,36,37] or even pre-trained neural networks designed to reduce high-dimensional data into simpler physical descriptors. Typically, in a GPBO driven experiment, a scalarizer function is predefined before the experiment based on prior knowledge or expectation of experiment outcomes; then, the same scalarizer function is applied throughout the entire GPBO driven experiment. However, a predefined scalarizer function often does not suffice for all scenarios due to the complexity of experimental data and lacks the flexibility for discovering unanticipated results. For example, a scalarizer used to identify the maximum peak intensity in spectroscopic data will fail to discern distinct spectra with peaks at different frequency and consequently will assign the identical scalar descriptor to peaks with different meanings (e.g. peaks with the same amplitude but at a different frequency). These can distort the true relationship between experiment conditions and target properties. Therefore, we need a quality control of the raw experiment results to check their compatibility with the predefined scalarizer function, in doing so, ensure the quality of the converted scalarizers and the training dataset.

To tackle the above challenges and limitations of scalarizer functions used in GPBO driven experimentation, we propose to use a $2^{nd}$ GP in tandem with the primary GPBO, as a solution to dynamically constrain the exploration space to areas that potentially produce more valuable results. This Dual-GP approach maintains the target property optimization driven by the traditional GPBO process and adds a $2^{nd}$ GP to dynamically constrain the experimental space based on observation of raw experimental data. The constraint of the $2^{nd}$ GP can be according to the compatibility between raw experimental data and scalarizer function, or the quality of the raw data, or additional assessment of material properties, etc. We also add an interface that allows human-in-the-loop intervention to account for unanticipated results. We demonstrate the application of the Dual GP in synthetic model data and experimental pulsed laser deposition (PLD) synthesis data; however, this approach can be applied to any other GPBO driven experiments as well.

## Results and Discussion

*The Dual-GP workflow*

Traditionally, as shown in Figure 1a, a single GP within a BO framework starts by assessing seed experiment conditions. The seed conditions are selected either randomly or by human choice. In this GPBO driven experiment loop, it is the scalar physical descriptor rather than raw experiment data that is used for GPBO analysis. Therefore, defining a scalarizer function, based on prior knowledge, is a crucial step for deriving physical descriptors from raw experimental data. As previously discussed, pre-defined scalarizer functions may fail to apply meaningful descriptors to certain data, resulting in irrelevant or meaningless scalar values that contaminate the dataset and mislead the GP approximation and BO selection. Failure of a scalarizer function can arise from various factors, such as large noise in the spectra, the presence of outliers, or the emergence of new physical phenomena not accounted for by the pre-defined scalarizer function. Additionally, it is virtually impossible to form a robust scalarizer to handle data that we cannot anticipate.

Figure 1b-g showcases the use of a scalarizer function to transform raw data into a scalar descriptor, illustrating the limitations of a predefined scalarizer function in analyzing experimental results. The raw spectral data displayed in Figure 1b-g are from the HybriD3 materials database,[38] which provides a collection of experimental and computational data on hybrid organic-inorganic compounds (HOI). These figures specifically present experimental photoluminescence (PL) spectroscopy results for HOI including hybrid organic-inorganic perovskites (HOIPs). An essential application of HOIPs lies in photovoltaic and light-emitting devices, where tuning the bandgap is crucial for either enhancing light absorption for photovoltaics or achieving emitting light of a specific color for light emitting devices. The bandgap can be inferred from the PL emission wavelength, thus it has been used as a valuable physical descriptor for optimizing HOIPs bandgap[35]. The PL emission wavelength can be extracted from the raw spectrum by identifying the PL peak position, which can be accomplished using the *find_peaks* function in *SciPy*.[39] This function allows users to customize parameters like peak height and width, and returns details of the identified peaks, including peak positions. Thus, we use this *find_peaks* function as a scalarizer function to convert PL raw spectrum to physical descriptor of emission wavelength, details regarding the analysis of these PL spectra and scalar descriptors conversion can be found in the Supplementary Notebook II that is also publicly available on GitHub.

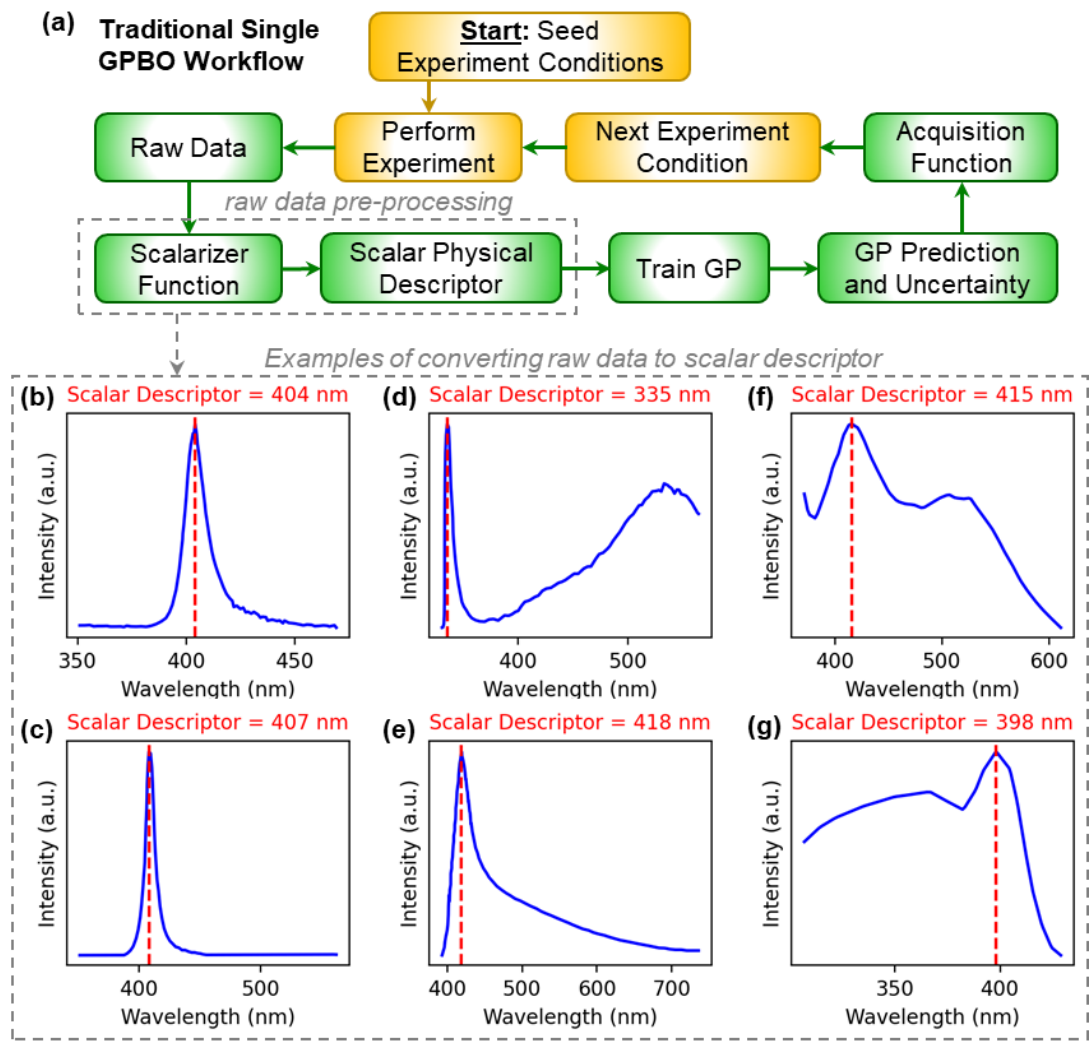

**Figure 1.** The workflow of GPBO driven experiments. (a), The workflow for traditional GPBO driven experiments starts with a few seed experimental conditions; performing experiments followed by preprocessing of converting the raw experimental data to scalar physical descriptors using a predefined scalarizer function; then the experiment conditions and scalar descriptors form a trainset to train a GP (surrogate model); next acquisition function determines the next experimental conditions based on the GP prediction and uncertainty. (b-g), examples of raw experimental photoluminescence (PL) results of hybrid organic inorganic compounds and corresponding scalar descriptors. The find_peaks function in SciPy is used as a scalarizer function to convert raw PL to scalar descriptor of emission wavelength. (b-c) shows scalar descriptors of high quality, while the scalar descriptors (that only represent the wavelength of the highest PL peak) in (d-g) miss crucial information in the raw spectra. This variation in scalarizer quality can mislead the GPBO driven experiments.

As indicated by the vertical red dashed lines in Figures 1b-g, the scalarizer function effectively identifies the wavelength of the highest peak. However, the quality of these scalarizers varies significantly. For instance, the scalarizers in Figures 1b and 1c are of high quality, where the raw PL spectra predominantly contain a single major peak.[40,41] In contrast, scalarizers from Figures 1d-f reveal significant limitations: the scalarizer only marks the highest emission peak, failing to account for additional phenomena in the spectra—e.g., a secondary broad peak appearing after 400 nm in Figure 1d,[41] significant asymmetry of the peak in Figure 1e,[42] and a secondary shoulder peak in Figure 1f and 1g.[43,44] These features, which involve additional physics like broad emission[41], self-trapped exciton[42], and ligand-contributed emissions[43,44], are critical for bandgap engineering but are overlooked by the predefined scalarizer function. Notably, it is virtually impossible to define a scalarizer function that can capture all known physical phenomena in the raw results, let alone unknown aspects.

Therefore, the quality of scalar physical descriptors can significantly vary due to complexities in the raw experimental data. Using these descriptors of varying quality in a training set could potentially mislead the GPBO driven experiment; for instance, although Figure 1e and 1f result in similar scalarizers (415 nm vs. 418 nm), the exact materials' properties, which can be examined from the raw spectra, are significantly different. To address this issue, we propose to use a 2$^{nd}$ GP as an observer, as shown in Figure 2, which assesses the quality of the raw data or its compatibility with the predefined scalarizer function to predict the applicability of the scalarizer function in the experimental space. This approximation can be used to assign a quality score to the experimental space, which examines relevance of the acquired raw data and the predefined scalarizer function. Subsequently the quality score can be used as a dynamic constraint on the exploration space, constraining the BO to focus on the space where the scalarizer is likely of high quality. This dynamic constraint has the potential to further accelerate the BO workflow by filtering out the space with low probability of acquiring useful data and low quality scalarizers.

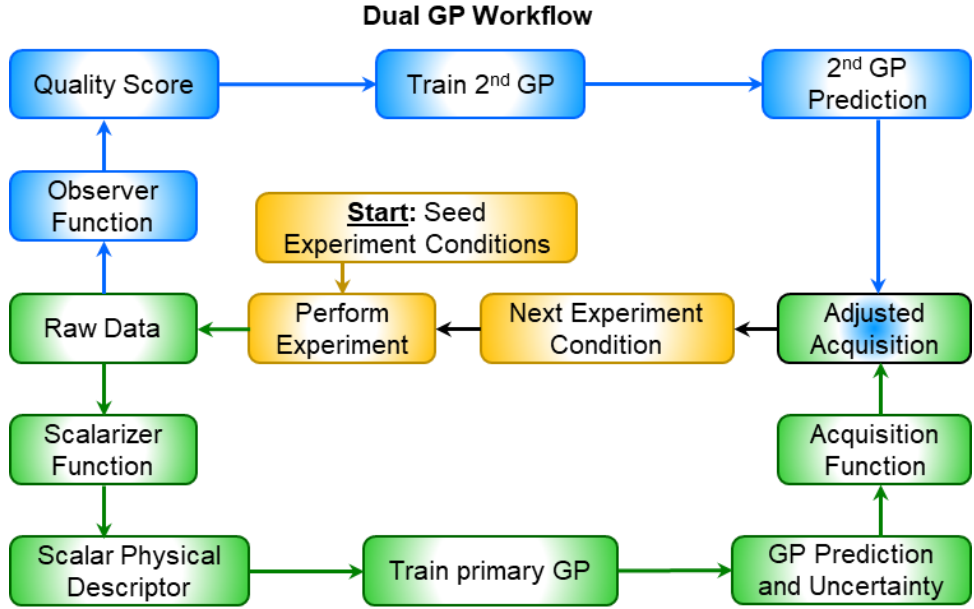

**Figure 2.** The workflow of a Dual-GP driven experiment. A 2nd GP analyzes the quality of resultant scalarizer, this quality can be obtained via examining the compatibility of the scalarizer function and the raw data. This prediction is used to actively refine the exploration space and modify the acquisition values. In doing so, the aim is to make the primary GPBO (which is driving experiment) avoid noisy or infeasible space.

We propose to design an observer function by comparing the on-the-fly raw experimental spectral data against a reference spectrum, this reference spectrum embodies the expected outcomes and is highly compatible with the predefined scalarizer function. The reference spectrum can be from seed experiments or theory. This comparison can be quantified by metrics such as Structural Similarity Index (SSI), Mean Square Error (MSE), etc., to assess whether the real time raw spectrum is compatible with the predefined scalarizer function. The 2nd GP is trained on the metric and refines the experimental space to focus on regions likely to yield spectra relevant to the predefined scalarizer function. This strategy dynamically adjusts the exploration space of the primary GPBO with insights from the 2nd GP, increasingly focusing on the space predicted to align with experimental goals. Thus, the integration of a 2nd GP enables a dynamic, goal-aligned refinement of the experimental landscape, ensuring a more streamlined and efficient exploration.

*Comparing single GPBO and Dual-GP BO*

To illustrate how Dual-GP can discern and prioritize the experimental space, we conducted a numerical experiment using a synthetic spectrum model with varying noise and compared the single GP and Dual-GP methods' ability to reconstruct the ground truth and determine the model's parameters based on limited observations. Each input $x$ produces a spectrum with a single Gaussian peak whose amplitude is given by equation 1:

$$y = (A + Bx)\sin(10x) + C \tag{1}$$

where A = 0.5; B = -1.2; and C = 0.5. We introduced higher noise to the spectra in the range x ∈[0.7, 1.0] to simulate "bad" experimental measurement conditions. Consequently, the amplitude extracted by the scalarizer function deviates from the true function within this range, as shown in Figure 3a, which can potentially mislead the GPBO optimization. A few examples of spectra are presented in Figure 3b. We implemented a structured GP (sGP)[45] for the primary GP in order to predict the model parameters. In sGP, we structured the mean function of the GP with the functional form of the amplitude model and placed a prior distribution on the parameters. We refer Ref [45] for further interest in sGP. The scalarizer function is constructed with *find_peaks* method to extract the peak amplitude as the scalar physical descriptor and the BO used an uncertainty-based acquisition function that selects the next point based on maximum GP uncertainty. The quality metric for 2nd GP training is quantified *via* SSI between measured spectra and a reference spectrum; this reference spectrum is an example of a low noise spectrum that has a good compatibility with the predefined scalarizer function and results in a scalarizer of high quality. The SSI of all synthetic spectra is in Supplementary Notebook II that is also publicly available in GitHub; high noise spectrum generally led to low SSI. The 2nd GP is trained by the SSI of measured spectra and predicts the SSI in the unexplored space. Thus, the predicted SSI of the unexplored space indicates the possibility of the unexplored space to produce high quality scalarizers. The acquisition function of the primary GPBO is set to zero where the predicted SSI score < 0.3, in doing so, a constraint is applied to the primary GPBO to only explore the space where the SSI score is larger than 0.3, which has a larger potential to produce high quality results and scalarizers.

Results for the single GP and Dual-GP are shown in Figure 3c and 3d, respectively, after 50 exploration steps. The single GPBO selected many points within the high noise subspace which reduce the GP surrogate's reconstruction accuracy, hence hindering the optimization process. In comparison, the 2nd GP in the Dual-GP effectively identified that the subspace with high noise is the region $x \in [0.7, 1.0]$, and hence ensures the primary GPBO avoids this subspace. Throughout exploration, the parameters of the structured mean function are updated to capture the underlying system behavior. Thus, by comparing the evolution of mean function parameters with the ground truth parameters, we can gain insights into the optimization process of single GP and Dual-GP. As shown in Figure 3e-g, the predicted parameters approach the ground truth quicker in Dual-GP driven optimization, indicating a more efficient optimization with Dual-GP.

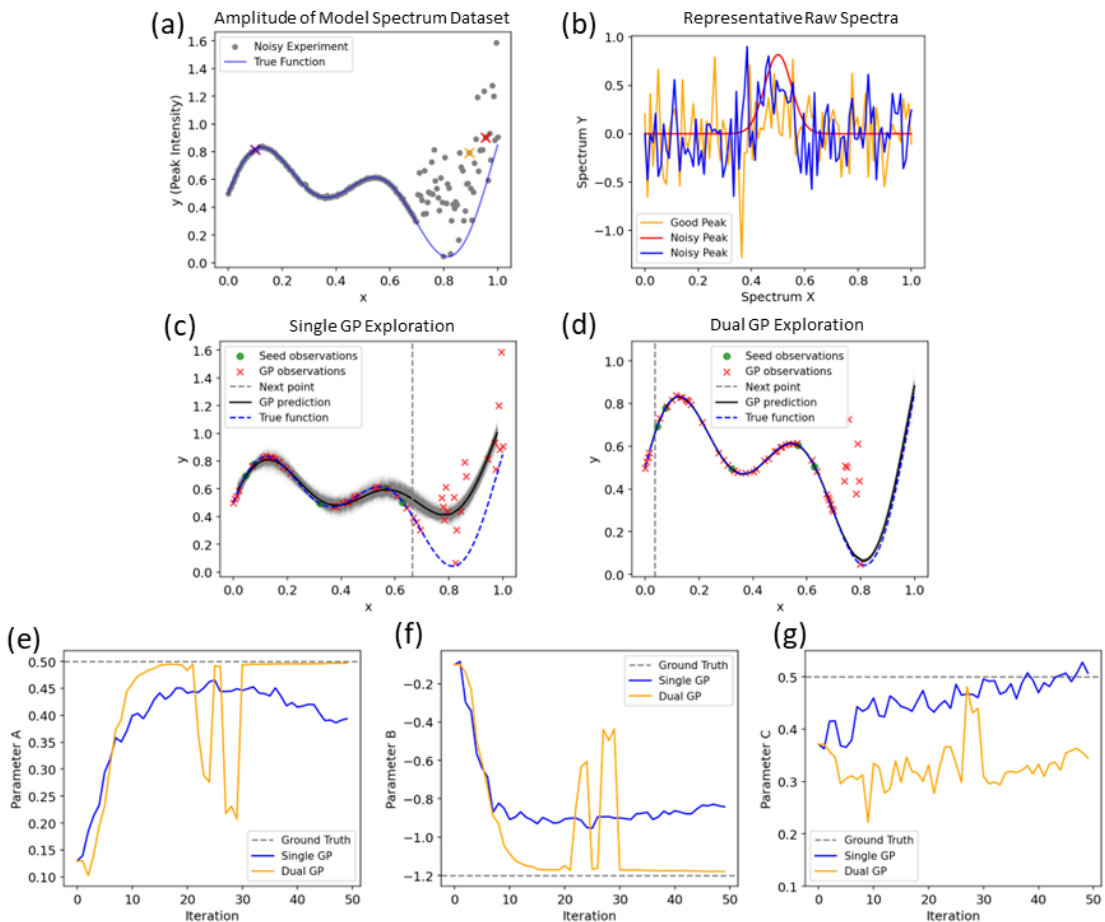

**Figure 3.** Dual-GP driven exploration in a simulated scenario where a sub-space contains high noise. (a) amplitude of the synthetic peak model data for the Dual-GP test, here the noise quickly increases when x > 0.7, leading to resultant scalarizers diverging from the true function. (b) Examples of a good raw spectrum and noisy spectra. (c) Single GP exploration result after 50 iterations, (d) Dual-GP exploration result after 50 iterations. (e-g) Show the sGP parameter prediction of Dual-GP approaches the true values more quickly than the single GP.

Above we used a predefined metric (i.e., SSI of raw experimental spectra to a reference spectrum) to assess the quality of raw experimental data and its compatibility with the predefined scalarizer function, this metric is defined according to our prior knowledge and/or anticipation of the experimental results. In some cases of real experiments, it is not possible to anticipate how the real-time experimental results will look like, and hence the quality of raw results cannot be assessed via a predefined metric; in these cases, real time human evaluation becomes an invaluable metric for determining the quality of the raw data and if it can yield meaningful scalarizers. To integrate human in the GPBO loop using the Dual-GP approach, human experts can review the collected raw spectral data and assign a quality score to each spectrum according to knowledge and/or interest. These scores, reflecting the relevance and the quality of the spectra, are then used to train the 2$^{nd}$ GP. Incorporating human assessment in this manner offers significant flexibility and adaptability, making it a universally applicable approach in scenarios where prior knowledge and reasonable anticipation about the material system is limited or unavailable.

To demonstrate how the human in loop Dual-GP model can prioritize the experimental domain, we generated another synthetic spectral dataset using equation (1) with peak amplitude parameters A = 0.3; B = -1; and C = 0.5 and, instead of increased noise as before, we introduced random distortions which alter the raw spectra in the region $x \in [0.3, 0.6]$, this random distortion to raw data is to mimic the 'unexpected' scenario in real experiment. The true amplitude function is shown in Figure 4a and the examples of distorted spectra are shown in Figure 4b. The assumption is that these distortions are unknown prior to experiment and cannot be reasonably represented by a predefined metric, necessitating real-time human assessment of the quality of raw spectra and their compatibility with the scalarizer function. We used the same sGP scheme and acquisition function as the previous experiment to demonstrate the human-in-loop Dual GP exploration of this model data. For the 2$^{nd}$ GP, after every 9 (this can be modified by users) exploration steps, the human expert is prompted to rank the last 9 spectra from 1-10 with 10 being "good" and this score was given to the 2$^{nd}$ GP to predict where high score spectra may be. Based on the prediction, the next iteration sampling is constrained in the subspace where the predicted score is >3, which is likely to lead to higher quality experimental data.

Results for the single GP and Dual-GP are shown in Figure 4c and 4d, respectively, after 50 exploration steps. Again, the single GPBO fails to reconstruct the true function in the distorted subspace. In contrast, the 2<sup>nd</sup> GP in the Dual-GP again identified that the distorted space is $x \in [0.3, 0.59]$, which is well aligned with the ground truth $x \in [0.3, 0.6]$. Subsequently, the Dual-GP workflow filtered out the acquired data from this subspace and constrained the exploration space to avoid sampling in this subspace. Through this approach, the Dual-GP effectively identifies valuable data, ensuring the primary GP focuses on high-quality spectra for more accurate and efficient optimization. A comparison of true function parameters predicted by the single GP and Dual-GP is shown in Figure 4e-g, with Dual-GP demonstrating better estimations of all three parameters, suggesting the potential of Dual-GP with human assessment for accelerated and efficient optimization.

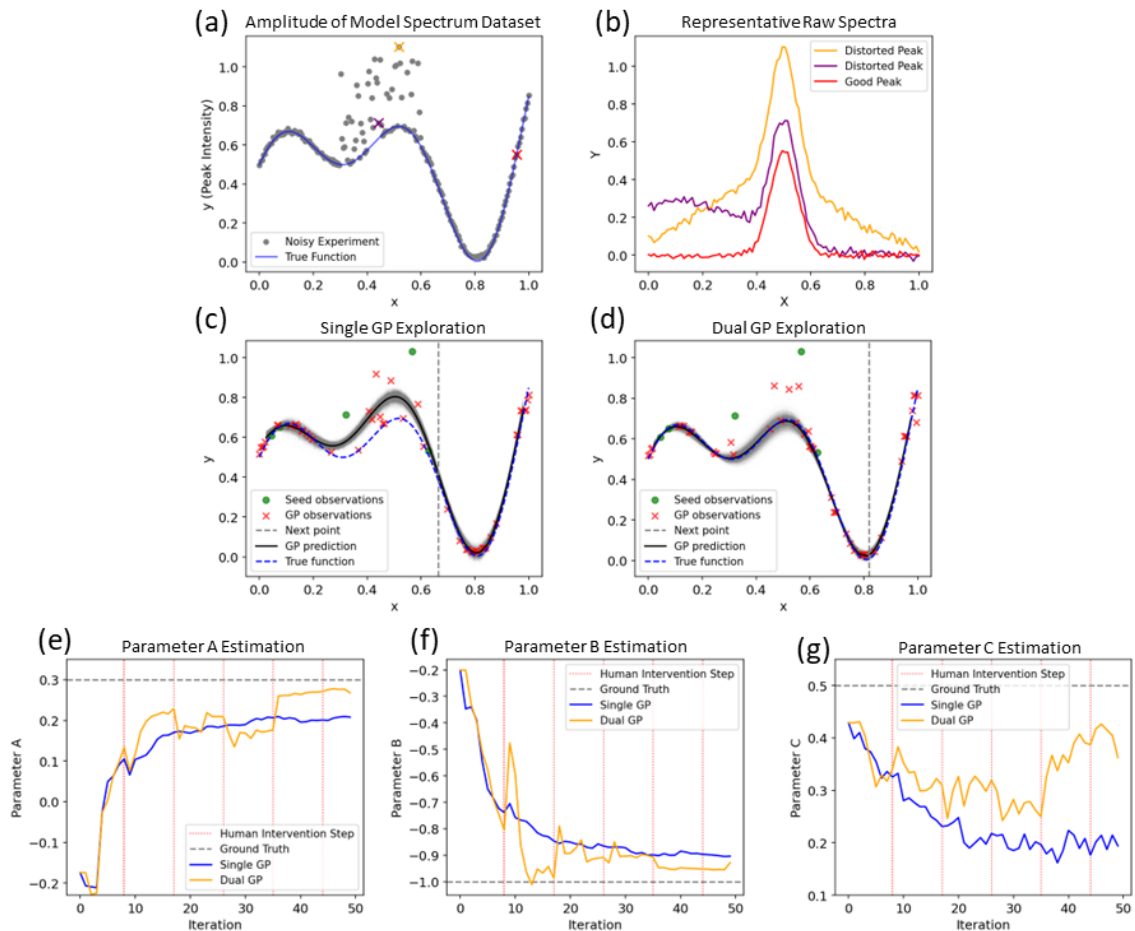

**Figure 4.** Human in the loop of Dual-GP exploration. (a) Amplitude of the synthetic peak model data, random distortion is added in the raw data in the space $x \in [0.3, 0.6]$, leading to resultant scalarizers diverging from the true function. (b) Examples of a good raw spectrum and distorted spectra. (c) Single GP exploration result after 50 iterations, (d) Dual-GP exploration result after 50 iterations, here human experts assess raw spectra every 9 iterations (this can be flexible) and assign

a quality score to the raw spectra based on the assessment; low quality score indicates the raw spectrum is not compatible with the predefined scalarizer function for various reasons. (e-g) Show that Dual-GP approximated the true function parameter better than the single GP.

*Human in the loop Dual-GP for Real Experimental Data*

After demonstrating the principle of Dual GP, we implemented this methodology in an autonomous PLD experiment data to assess its effectiveness for real world application. The full details of the autonomous PLD experiment can be found in our previous work[22]. Briefly, $WSe_2$ thin films of nominally monolayer thickness were grown by PLD using co-ablation of $WSe_2$ and Se targets varying 4 growth parameters: pressure ($P$), substrate temperature ($T$), and laser energy on the $WSe_2$ and Se targets $E_1$ and $E_2$, respectively. The scalarizer function that was used was derived from the Raman spectrum of each film after growth and calculated as the ratio of the primary $WSe_2$ $E_{2g}$+$A_{1g}$ Raman peak height and width – a high "score" is achieved from intense, narrow peaks. Traditional GPBO was used to explore the 4D parameter space to maximize the Raman score using the expected improvement (EI) acquisition function. While this experiment was successful, the GPBO directed the synthesis of many films in regions of the parameter space that continually produced poor quality samples. This is because a high dimensional space populated with only 10s of samples results in high GP uncertainty and the BO tended to favor exploration. Because the growth window was narrow in at least 1 of the parameters, prolonged exploration led to numerous poor-quality samples. This effect is expected in traditional GPBO but when the maximum budget for total samples is small, as it is with PLD experiments, it is highly undesirable. Further, pure optimization is not always of interest for synthesis science. Experimentalists usually want to understand the synthesis response surface to determine mechanisms of film growth. In this case, uncertainty-based acquisition is desired to achieve a representative GP surrogate, but human guidance is needed to prevent frivolous exploration.

In the Dual-GP PLD experiment, we used the final GP surrogate from the initial study to act as the "experimental ground truth" to evaluate the reconstruction error from uncertainty-based exploration (UE) using single GPBO, Dual-GP, and random sampling. For the quality score of raw data in Dual-GP, we constructed it by ranking 108 Raman spectra from the experiment to make approximation with a GP. During the human in the loop Dual-GP experiment, the quality score was sampled to train the $2^{nd}$ GP. With UE, the primary GPBO selects the next point based on maximum uncertainty. For each scenario, the same 10 samples were used as seed points. In the

Dual-GP case, the exploration is constrained in the subspace where the quality score is >7 for the first 50 steps and > 5 after that. The parameter space was discretized into 15x15x15x15 to have 50625 possible combinations of parameters. Each experiment was run for 200 steps, sampling 0.4% of the total space.

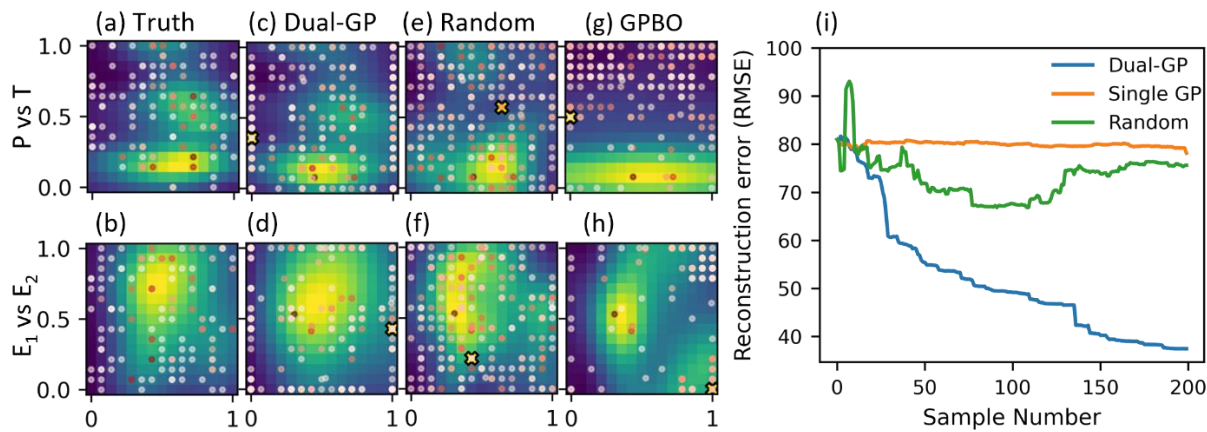

**Figure 5.** Simulated Dual-GP, human in the loop, uncertainty-based exploration in 4D space using experimental data from an autonomous PLD experiment. (a-b) The ground truth response surface projected into the *P vs. T* and *E₁ vs. E₂* planes. (c-d) The Dual-GP reconstruction closely matches the ground truth. (e-f) Random sampling performs better than (e-f) pure uncertainty exploration with traditional GPBO. The points in each map represent sampled points, where red color indicates a high score. The axes of each map are normalized. (i) The root mean squared error of ground truth reconstruction vs sample number for all three cases shows that human in the loop Dual-GP quickly outperforms random sampling and traditional GPBO when using uncertainty-based exploration, which is attractive for synthesis science applications where pure optimization is not the goal of the experiment.

Figure 5a-b show the experimental ground truth of the Raman score response surface projected into the *P vs. T* and *E₁ vs. E₂* planes, respectively. The Dual-GP reconstruction closely matches the ground truth and sampled the space efficiently to reduce the root mean squared error (RMSE) of reconstruction rapidly within the first 50 steps (Figure 5i). Random sampling performed the next best, but still poorly reconstructs the ground truth. Lastly, traditional GPBO completely failed to reconstruct the ground truth with very little improvement in the RMSE over all 200 steps. The high uncertainty in the sparsely sampled space leads to many samples at the edges in the *E₁ vs E₂* plane (Figure 5h) and sampled the high-pressure region almost exclusively. It should be noted that the goal of this synthesis simulation was not to locate the maximum as quickly as possible but rather to quickly build an effective surrogate model for the synthesis space

with sparse sampling. The role of the human in this scenario is to assess the raw experimental data, this assessment can be used to dynamically adjust the feasible space while still allowing for uncertainty-based exploration. These results indicate that Dual-GP with human assessment can lead to more efficient optimization in experiments.

**Conclusions**

In summary, we demonstrate that the Dual-GP approach represents an advancement in GPBO driven autonomous experimentation, addressing a key limitation inherent to GPBO applications in real world experiments. By introducing a $2^{nd}$ GP to dynamically constrain the experimental space based on the observation of raw experimental results, the Dual-GP approach mitigates the shortcomings of traditional GPBO optimization and enhances the adaptability and accuracy of the optimization process. This approach effectively isolates more promising experimental spaces for BO sampling and improves the quality of obtained data. Furthermore, we also developed a strategy for human-in-the-loop Dual GP optimization, allowing experts to assess and adjust experiments, ensuring that unanticipated scenarios in real world experiments are appropriately managed. It has also been shown that similar human-AI collaboration improves semiconductor process development efficiency[46]. The demonstrated application of the Dual-GP approach in both synthetic and real-world experimental settings indicate its potential in broad autonomous experiments across various domains. For materials developments that are expensive, time consuming, and difficult to automate, the Dual-GP approaches is an effective technique to rapidly understand quantitative trends of material properties vs. experimental conditions in large parameter spaces with a minimal number of samples by effectively infusing human expertise into the autonomous workflow. The Dual-GP approach can also be used to incorporate on-the-fly experimentation in autonomous platforms, offline in-depth investigation, and theory, etc., enabling cross-facilities, asynchronous co-optimization for accelerated research.[47]

**Acknowledgements**

This research was supported by the Center for Nanophase Materials Sciences (CNMS), which is a US Department of Energy, Office of Science User Facility at Oak Ridge National Laboratory.

**Conflict of Interest**

The authors declare no conflict of interest.

**Authors Contribution**

Y.L. conceived the idea. Y.L. performed the investigation. S.H. performed PLD experiment and data analysis. All authors contributed to discussions and manuscript editing.

**Data Availability**

The data and code of this study is provided in the https://github.com/yongtaoliu/dual-GP.git. The approach is built using GPax https://gpax.readthedocs.io/en/latest/.